\documentclass[sigconf,review=false,screen=false,authorversion=true,anonymous=false]{acmart}

\usepackage{diagbox}
\usepackage{subcaption}
\usepackage{multirow}
\usepackage{balance} 
\AtBeginDocument{%
  \providecommand\BibTeX{{%
    \normalfont B\kern-0.5em{\scshape i\kern-0.25em b}\kern-0.8em\TeX}}}

\copyrightyear{2020} 
\acmYear{2020} 
\setcopyright{acmlicensed}\acmConference[MM '20]{Proceedings of the 28th ACM International Conference on Multimedia}{October 12--16, 2020}{Seattle, WA, USA}
\acmBooktitle{Proceedings of the 28th ACM International Conference on Multimedia (MM '20), October 12--16, 2020, Seattle, WA, USA}
\acmPrice{15.00}
\acmDOI{10.1145/3394171.3413740}
\acmISBN{978-1-4503-7988-5/20/10}

\begin{document}
\fancyhead{}
\title{FastLR: Non-Autoregressive Lipreading Model  \\  with Integrate-and-Fire}

\author{Jinglin Liu$^*$}
\affiliation{
\institution{Zhejiang University}
}
\email{jinglinliu@zju.edu.cn}

\author{Yi Ren$^*$}
\affiliation{
\institution{Zhejiang University}
}
\email{rayeren@zju.edu.cn}

\author{Zhou Zhao$\dagger$}
\affiliation{
\institution{Zhejiang University}
}
\email{zhaozhou@zju.edu.cn}

\author{Chen Zhang}
\affiliation{
\institution{Zhejiang University}
}
\email{zc99@zju.edu.cn}

\author{Baoxing Huai}
\affiliation{
\institution{HUAWEI TECHNOLOGIES CO., LTD.}
}
\email{huaibaoxing@huawei.com}

\author{Nicholas Jing Yuan}
\affiliation{
\institution{Huawei Cloud BU}
}
\email{nicholas.yuan@huawei.com}

\begin{abstract}
Lipreading is an impressive technique and there has been a definite improvement of accuracy in recent years. However, existing methods for lipreading mainly build on autoregressive (AR) model, which generate target tokens one by one and suffer from high inference latency. To breakthrough this constraint, we propose FastLR, a non-autoregressive (NAR) lipreading model which generates all target tokens simultaneously. NAR lipreading is a challenging task that has many difficulties: 1) the discrepancy of sequence lengths between source and target makes it difficult to estimate the length of the output sequence; 2) the conditionally independent behavior of NAR generation lacks the correlation across time which leads to a poor approximation of target distribution; 3) the feature representation ability of encoder can be weak due to lack of effective alignment mechanism; and 4) the removal of AR language model exacerbates the inherent ambiguity problem of lipreading. Thus, in this paper, we introduce three methods to reduce the gap between FastLR and AR model: 1) to address challenges 1 and 2, we leverage integrate-and-fire (I\&F) module to model the correspondence between source video frames and output text sequence. 2) To tackle challenge 3, we add an auxiliary connectionist temporal classification (CTC) decoder to the top of the encoder and optimize it with extra CTC loss. We also add an auxiliary autoregressive decoder to help the feature extraction of encoder. 3) To overcome challenge 4, we propose a novel Noisy Parallel Decoding (NPD) for I\&F and bring Byte-Pair Encoding (BPE) into lipreading. Our experiments exhibit that FastLR achieves the speedup up to 10.97$\times$ comparing with state-of-the-art lipreading model with slight WER absolute increase of 1.5\% and 5.5\% on GRID and LRS2 lipreading datasets respectively, which demonstrates the effectiveness of our proposed method.\footnote{$^*$ Equal contribution. $\dagger$ Corresponding author.}
\end{abstract}

\ccsdesc[500]{Computing methodologies~Computer vision}
\ccsdesc[500]{Computing methodologies~Speech recognition}

\keywords{Lip Reading; Non-autoregressive generation; Deep Learning}

\settopmatter{printacmref=false, printfolios=false}

\maketitle

{\fontsize{8pt}{8pt} \selectfont
\textbf{ACM Reference Format:}\\
Jinglin Liu, Yi Ren, Zhou Zhao, Chen Zhang, Baoxing Huai, and Jing Yuan. 2020. FastLR: Non-Autoregressive Lipreading Model with Integrate-and- Fire. In \text{\it Proceedings of the 28th ACM International Conference on Multimedia}  \text{\it (MM’20), October 12--16, 2020, Seattle, WA, USA.} ACM, New York, NY, USA, 9 pages. https://doi.org/10.1145/3394171.3413740 }

\section{Introduction}
Lipreading aims to recognize sentences being spoken by a talking face, which is widely used now in many scenarios including dictating instructions or messages in a noisy environment, transcribing archival silent films, resolving multi-talker speech~\cite{afouras2018deepav} and understanding dialogue from surveillance videos. However, it is widely considered a challenging task and even experienced human lipreaders cannot master it perfectly~\cite{assael2016lipnet, shillingford2018large}. Thanks to the rapid development of deep learning in recent years, there has been a line of works studying lipreading and salient achievements have been made.

Existing state-of-the-art methods mainly adopt autoregressive (AR) model, either based on RNN~\cite{zhang2019spatio,zhao2019hearing}, or Transformer~\cite{afouras2018compare,afouras2018deepav}. Those systems generate each target token conditioned on the sequence of tokens generated previously, which hinders the parallelizability. Thus, they all without exception suffer from high inference latency, especially when dealing with the massive videos data containing hundreds of hours (like long films and surveillance videos) or real-time applications such as dictating messages in a noisy environment. 

To tackle the low parallelizability problem due to AR generation, many non-autoregressive (NAR) models~\cite{gu2017non, lee2018deterministic,guo2019non,wang2019non,ma2019flowseq,liu2020task,ren2020study} have been proposed in the machine translation field. The most typical one is NAT-FT~\cite{gu2017non}, which modifies the Transformer~\cite{vaswani2017attention} by adding a fertility module to predict the number of words in the target sequence aligned to each source word. Besides NAR translation, many researchers bring NAR generation into other sequence-to-sequence tasks, such as video caption~\cite{ren2019fastspeech,ren2020fastspeech}, speech recognition~\cite{chen2019non} and speech synthesis\cite{oord2017parallel, ren2019fastspeech}. These works focus on generating the target sequence in parallel and mostly achieve more than an order of magnitude lower inference latency than their corresponding AR models.

However, it is very challenging to generate the whole target sequence simultaneously in lipreading task in following aspects:
\begin{itemize}
\item The considerable discrepancy of sequence length between the input video frames and the target text tokens makes it difficult to estimate the length of the output sequence or to define a proper decoder input during the inference stage. This is different from machine translation model, which can even simply adopt the way of uniformly mapping the source word embedding as the decoder input \cite{wang2019non} due to the analogous text sequence length.
\item The true target sequence distributions show a strong correlation across time, but the NAR model usually generates target tokens conditionally independent of each other. This is a poor approximation and may generate repeated words. \citet{gu2017non} terms the problem as "multimodal-problem".
\item The feature representation ability of encoder could be weak when just training the raw NAR model due to lack of effective alignment mechanism.
\item The removal of the autoregressive decoder, which usually acts as a language model, makes the model much more difficult to tackle the inherent ambiguity problem in lipreading.
\end{itemize}

In our work, we propose FastLR, a non-autoregressive lipreading model based on Transformer. To handle the challenges mentioned above and reduce the gap between FastLR and AR model, we introduce three methods as follows: 
\begin{itemize}
    \item To estimate the length of the output sequence and alleviates the problem of time correlation in target sequence, we leverage integrate-and-fire (I\&F) module to encoding the continuous video signal into discrete token embeddings by locating the acoustic boundary, which is inspired by \citet{dong2019cif}. These discrete embeddings retain the timing information and correspond to the target tokens directly.
    \item To enhance the feature representation ability of encoder, we add the connectionist temporal classification (CTC) decoder on the top of encoder and optimize it with CTC loss, which could force monotonic alignments. Besides, we add an auxiliary AR decoder during training to facilitate the feature extraction ability of encoder.
    \item To tackle the inherent ambiguity problem and reduce the spelling errors in NAR inference, we first propose a novel Noisy Parallel Decoding (NPD) for I\&F method. The rescoring method in NPD takes advantages of the language model in the well-trained AR lipreading teacher without harming the parallelizability. Then we bring Byte-Pair Encoding (BPE) into lipreading, which compresses the target sequence and makes each token contain more language information to reduce the dependency among tokens compared with character level encoding.
\end{itemize}

The core contribution of this work is that, we propose a non-autoregressive lipreading system, and present several elaborate methods metioned above to bridge the gap between FastLR and state-of-the-art autoregressive lipreading models.

The experimental results show that FastLR achieves the speedup up to 10.97$\times$ comparing with state-of-the-art lipreading model with slight WER increase of 1.5\% and 5.5\% on GRID and LRS2 lipreading datasets respectively, which demonstrates the effectiveness of our proposed method. We also conduct ablation experiments to verify the significance of all proposed methods in FastLR.

\section{Related Works}
\subsection{Deep Lipreading}
Prior works utilize deep learning for lipreading. The first typical approach is LipNet \cite{assael2016lipnet} based on CTC~\cite{graves2006ctc}, which takes the advantage of the spatio-temporal convolutional front-end feature generator and GRU \cite{chung2014gru}. Further, \citet{stafylakis2017combining} propose a network combining the modified 3D/2D-ResNet architecture with LSTM. \citet{afouras2018deepav} introduce the Transformer self-attention architecture into lipreading, and build TM-seq2seq and TM-CTC. The former surpasses the performance of all previous work on LRS2-BBC dataset by a large margin. To boost the performance of lipreading, \citet{ctchyp2018} present a hybrid CTC/Attention architecture aiming to obtain the better alignment than attention-only mechanism, \citet{zhao2019hearing} provide the idea that transferring knowledge from audio-speech recognition model to lipreading model by distillation. 

However, state-of-the-art methods, either based on recurrent neural network ~\cite{zhang2019spatio,zhao2019hearing} or Transformer~\cite{afouras2018compare,afouras2018deepav}, take in the input video sequence and generates the tokens of target sentence $y$ in a recurrent structure during the inference process. And they all suffer from the high latency.
\subsection{Non-Autoregressive Decoding}
An autoregressive model takes in a source sequence $x = (x_1, x_2, ..., \\ x_{T_x})$ and then generates words in target sentence $y = (y_1, y_2, ..., y_{T_y})$ one by one with the causal structure during the inference process \cite{sutskever2014sequence,vaswani2017attention}.
To reduce the inference latency, \citet{gu2017non} introduce non-autoregressive model based on Transformer into the machine translation field, which generates all target words in parallel. The conditional probability can be defined as
\begin{equation}
P(y|x) = P(T_y|x)\prod_{t=1}^{T_y}P(y_t|x) ,
\end{equation}
where $T_y$ is the length of the target sequence gained from the fertility prediction function conditioned on the source sentence. Due to the multimodality problem \cite{gu2017non}, the performance of NAR model is usually inferior to AR model. Recently, a line of works aiming to bridge the performance gap between NAR and AR model for translation task has been presented \cite{ghazvininejad2019mask,guo2019non}. 

Besides the study of NAR translation, many works bring NAR model into other sequence-to-sequence tasks, such as video caption \cite{yang2019non}, speech recognition \cite{chen2019non} and speech synthesis \cite{oord2017parallel,ren2019fastspeech}. 

\begin{figure*}[htb]
  \centering
  \includegraphics[width=0.92\linewidth]{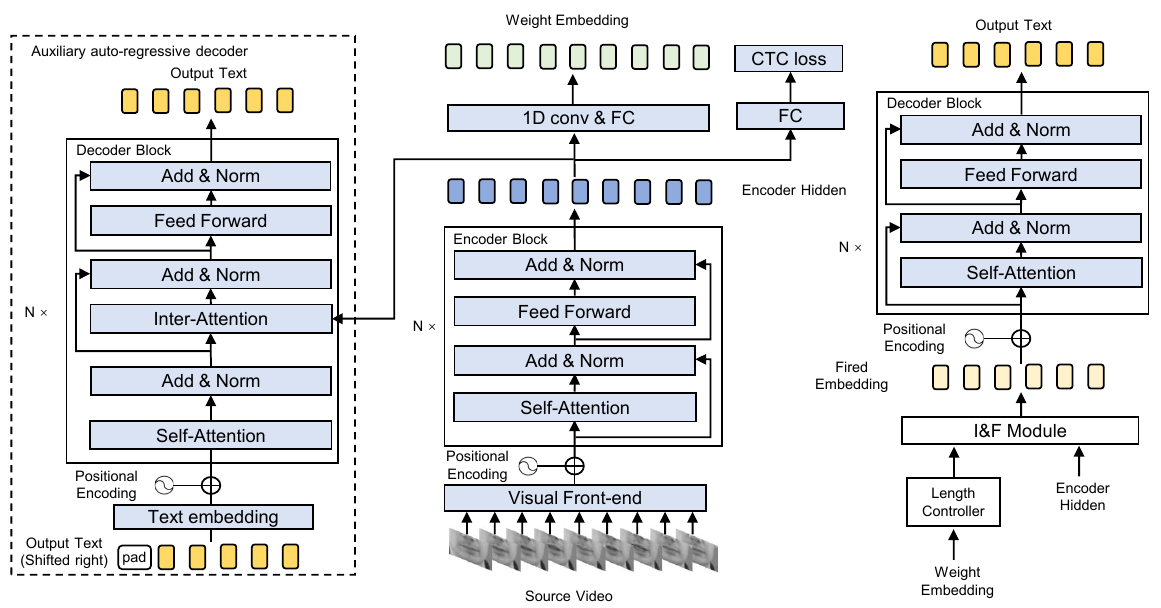}
  \caption{The overview of the model architecture for FastLR.}
  \label{archi_model}
\end{figure*}

\subsection{Spike Neural Network}
The integrate-and-fire neuron model describes the membrane potential of a neuron according to the synaptic inputs and the injected current \cite{burkitt2006review}. It is bio-logical and widely used in spiking neural networks. Concretely, the neuron integrates the input signal forwardly and increases the membrane potential. Once the membrane potential reaches a threshold, a spike signal is generated, which means an event takes place. Henceforth, the membrane potential is reset and then grows in response to the subsequent input signal again. It enables the encoding from continuous signal sequences to discrete signal sequences, while retaining the timing information.

Recently, \citet{dong2019cif} introduce the integrate-and-fire model into speech recognition task. They use continuous functions that support back-propagation to simulate the process of integrate-and-fire. In this work, the fired spike represents the event that locates an acoustic boundary.

\section{Methods}
In this section, we introduce FastLR and describe our methods thoroughly. As shown in Figure~\ref{archi_model}, FastLR is composed of a spatio-temporal convolutional neural network for video feature extraction (visual front-end) and a sequence processing model (main model) based on Transformer with an enhenced encoder, a non-autoregressive decoder and a I\&F module. To further tackle the challenges in non-autoregressive lipreading, we propose the NPD method for I\&F and bring byte-pair encoding into our method. The details of our model and methods are described in the following subsections\footnote{We introduce the visual front-end in section~\ref{para:visual_feature_ext} as it varies from one dataset to another.}:

\subsection{Enhenced Encoder}
\label{para:encoder_enhance}
The encoder of FastLR is composed of stacked self-attention and feed-forward layers, which are the same as those in Transformer~\cite{vaswani2017attention} and autoregressive lipreading model (TM-seq2seq\cite{afouras2018deepav}). Thus, we add an auxiliary autoregressive decoder, shown
in the left panel of Figure \ref{archi_model}, and by doing so, we can optimize the AR lipreading task with FastLR together with one shared the encoder during training stage. This transfers knowledge from the AR model to FastLR which facilitates the optimization.
Besides, we add the connectionist temporal classification (CTC) decoder with CTC loss on the encoder for forcing monotonic alignments, which is a widely used technique in speech recognition field. 
Both adjustments improve the feature representation ability of our encoder.

\subsection{Integrate-and-fire module}
\label{para:cif}
\begin{figure}[htb]
  \centering
  \includegraphics[width=\linewidth]{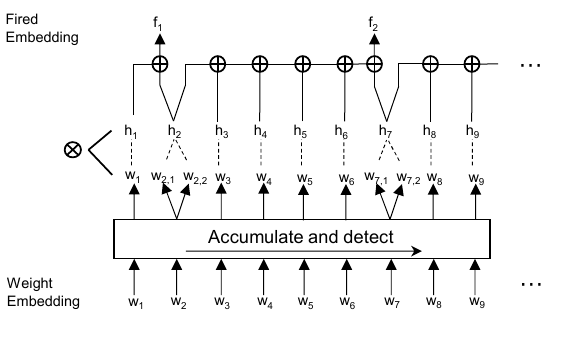}
  \caption{An example to illustrate how I\&F module works. $h$ respresents the encoder output hidden sequence. 
  In this case  $f_1 =  w_1 \cdot h_1 + w_{2,1} \cdot h_2, f_2 = w_{2,2} \cdot h_2 + w_3 \cdot h_3 + w_4 \cdot h_4 + w_5 \cdot h_5 + w_6 \cdot h_6 + w_{7,1} \cdot h_7$.}
  \label{cif_module}
\end{figure}

To estimate the length of the output sequence and alleviate the problem of time correlation in target sequence, we adopt continuous integrate-and-Fire (I\&F)~\cite{dong2019cif} module for FastLR. This is a soft and monotonic alignment which can be employed in the encoder-decoder sequence processing model. 
First, the encoder output hidden sequence $h = (h_1, h_2, …, h_m)$ will be fed to a 1-dimensional convolutional layer followed by a fully connected layer with sigmoid activation function. Then we obtain the weight embedding sequence $w = (w_1, w_2, …, w_m)$ which represents the weight of information carried in $h$. Second, the I\&F module scans $w$ and accumulates them from left to right until the sum reaches the threshold (we set it to 1.0), which means an acoustic boundary is detected. Third, I\&F divides $w_i$ at this point into two part: $w_{i,1}$ and $w_{i,2}$. $w_{i,1}$ is used for fulfilling the integration of current embedding $f_j$ to be fired, while $w_{i,2}$ is used for the next integration of $f_{j+1}$. Then, I\&F resets the accumulation and continues to scan the rest of $w$ which begins with $w_{i,2}$ for the next integration. This procedure is noted as "accumulate and detect". Finally, I\&F multiplies all $w_k$ (or $w_{k,1}, w_{k,2}$) in $w$ by corresponding $h_k$ and integrates them according to detected boundaries. An example is shown in Figure~\ref{cif_module}.

\subsection{Non-autoregressive Decoder}
Different from Transformer decoder, the self-attention of FastLR’s decoder can attend to the entire sequence for the conditionally independent property of NAR model. And we remove the inter-attention mechanism since FastLR already has an alignment mechanism (I\&F) between source and target. The decoder takes in the fired embedding sequence of I\&F $f = (f_1, f_2, …, f_n)$ and generates the text tokens $y = (y_1, y_2, … y_n)$ in parallel during either training or inference stage.

\subsection{Noisy parallel decoding (NPD) for I\&F}
\label{sec:npd_for_cif}
The absence of AR decoding procedure makes the model much more difficult to tackle the inherent ambiguity problem in lipreading. So, we design a novel NPD for I\&F method to leverage the language information in well-trained AR lipreading model.

In section \ref{para:cif}, it is not hard to find that, $\left \lfloor S \right \rfloor$ represents the length of predicted sequence $f$ (or $y$), where $S$ is the total sum of $w$. And \citet{dong2019cif} propose a scaling strategy which multiplies $w$ by a scalar $\frac{\widetilde{S}}{\sum_{i=1}^m w_i}$ to generate $w' = (w_1’, w_2’, …, w_m’) $, where $\widetilde{S}$ is the length of target label $\widetilde{y}$. By doing so, the total sum of $w’$ is equal to $\widetilde{S}$ and this teacher-forces I\&F to predict $f$ with the true length of $\widetilde{S}$ which would benefit the cross-entropy training. 

However, we do not stop at this point. Besides training, we also scale $w$ during the inference stage to generate multiple candidates of weight embedding with different length bias $\widetilde{b}$. When set the beam size $B = 4$, 
\begin{equation}
w’_{\widetilde{b}} = \frac{\sum_{i=1}^m w_i + \widetilde{b}}{\sum_{i=1}^m w_i} \cdot w \text{, where } \widetilde{b} \in [-4, 4] \cap \mathbb{Z} , 
\end{equation}
where $w = (w_1, w_2, …, w_m)$ is the output of I\&F module during inference and length bias $\widetilde{b}$ is provided in "Length Controller" module in Figure \ref{archi_model}.
Then, we utilize the re-scoring method used in Noisy Parallel Decoding (NPD), which is a common practice in non-autoregressive neural machine translation, to select the best sequence from these $2*B$ candidates via an AR lipreading teacher:
\begin{equation}
    w_{NPD} = \mathop{argmax}\limits_{w’_{\widetilde{b}}} p_{AR}(G(x, w’_{\widetilde{b}};\theta)|x;\theta) , 
\end{equation}
where $p_{AR}(A)$ is the probability of the sequence $A$ generated by autoregressive model; The $G(x, w;\theta)$ means the optimal generation of FastLR given a source sentence $x$ and weight embedding $w$, $\theta$ represents the parameters of model.

The selection process could leverage information in the language model (decoder) of the well-trained autoregressive lipreading teacher, which alleviates the ambiguity problem and gives a chance to adjust the weight embedding generated by I\&F module for predicting a better sequence length. Note that these candidates can be computed independently, which won’t hurt the parallelizability (only doubles the latency due to the selection process). The experiments demonstrate that the re-scored sequence is more accurate. 

\subsection{Byte-Pair Encoding}
Byte-Pair Encoding \cite{sennrich2015neural} is widely used in NMT \cite{vaswani2017attention} and ASR \cite{dong2019cif} fields, but rare in lipreading tasks.
BPE could make each token contain more language information and reduce the dependency among tokens compared with character level encoding, which alleviate the problems of non-autoregressive generation discussed before. In this work, we tokenize the sentence with moses tokenizer \footnote{https://github.com/moses-smt/mosesdecoder/blob/master/scripts/tokenizer\\/tokenizer.perl} and then use BPE algorithm to segment each target word into sub-words.

\subsection{Training of FastLR}
We optimize the CTC decoder with CTC loss. CTC introduces a set of intermediate representation path $\phi(y)$ termed as CTC paths for one target text sequence $y$. Each CTC path is composed of scattered target text tokens and blanks which can reduce to the target text sequence by removing the repeated words and blanks. The likelihood of $y$ could be calculated as the sum of probabilities of all CTC paths corresponding to it:
\begin{equation}
    P_{ctc}(y|x) = \sum\limits_{c \in \phi(y)} P_{ctc}(c|x)
\end{equation}
Thus, CTC loss can be formulated as:
\begin{equation}
\mathcal{L}_{ctc} = - \sum\limits_{(x,y)\in (\mathcal{X}\times\mathcal{Y})} \sum\limits_{c \in \phi(y)}P_{ctc}(c|x)
\end{equation}
where $(\mathcal{X}\times\mathcal{Y})$ denotes the set of source video and target text sequence pairs in one batch.

We optimize the auxiliary autoregressive task with cross-entropy loss, which can be formulated as:
\begin{equation}
    \mathcal{L}_{AR} = - \sum\limits_{(x,y)\in (\mathcal{X}\times\mathcal{Y})} log P_{AR}(y|x)
\end{equation}

And most importantly, we optimize the main task FastLR with cross-entropy loss and sequence length loss:
\begin{equation}
    \mathcal{L}_{FLR} = - \sum\limits_{(x,y)\in (\mathcal{X}\times\mathcal{Y})} \left [ log P_{FLR}(y|x) + (\widetilde{S_x} - S_x)^2 \right ]
\end{equation}
where the $\widetilde{S}$ and $S$ are defined in section \ref{sec:npd_for_cif}.

Then, the total loss function for training our model is:
\begin{equation}
    \mathcal{L} = \lambda_1\mathcal{L}_{ctc} + \lambda_2\mathcal{L}_{AR} + \lambda_3\mathcal{L}_{FLR}
\end{equation}
where the $\lambda_1$, $\lambda_2$, $\lambda_3$ are hyperparameters to trade off
the three losses.

\section{Experiments and Results}
\subsection{Datasets}
\paragraph{GRID}
The GRID dataset~\cite{griddataset} consists of 34 subjects, and each of them utters 1,000 phrases. It is a clean dataset and easy to learn. We adopt the split the same with \citet{assael2016lipnet}, where 255 random sentences from each speaker are selected for evaluation. In order to better recognize lip movements, we transform the image into gray scale, and crop the video images to a fixed $100\times50$ size containing the mouth region with Dlib face detector. Since the vocabulary size of GRID datasets is quite small and most words are simple, we do not apply Byte-Pair Encoding~\cite{sennrich2015neural} on GRID, and just encode the target sequence at the character level.
\paragraph{LRS2}
The LRS2 dataset contains sentences of up to 100 characters from BBC videos \cite{afouras2018compare}, which have a range of viewpoints from frontal to profile. We adopt the origin split of LRS2 for \text{train/dev/test} sets, which contains 46k, 1,082 and 1,243 sentences respectively. And we make use of the pre-train dataset provided by LRS2 which contains 96k sentences for pretraining. Following previous works \cite{afouras2018compare,afouras2018deepav,zhao2019hearing}, the input video frames are converted to grey scale and centrally cropped into $114\times114$ images. As for the text sentence, we split each word token into subwords using BPE~\cite{sennrich2015neural}, and set the vocabulary size to 1k considering the vocabulary size of LRS2. 

The statistics of both datasets are listed in Table~\ref{dataset_statistic}.

\begin{table}[htb]
  \caption{The statistics on GRID and LRS2 lip reading datasets. \textbf{Utt:} Utterance.}
  \begin{tabular}{c c c c c}
    \toprule
    Dataset & Utt. & Word inst. & Vocab & hours \\
    \midrule
    GRID  & 33k  & 165k & 51 & 27.5      \\
    \midrule
    LRS2 (Train-dev)  & 47k  & 337k & 18k & 29  \\
    \bottomrule
  \end{tabular}
  \label{dataset_statistic}
\end{table}

\subsection{Visual feature extraction}
\label{para:visual_feature_ext}
For GRID datasets, we use spatio-temporal CNN to extract visual features follow~\citet{3dmatch}. The visual front-end network is composed of four 3D convolution layers with 3D max pooling and RELU, and two fully connected layers. The kernel size of 3D convolution and pooling is $3 \times 3$, the hidden sizes of fully connected layer as well as output dense layer are both 256. We directly train this visual front-end together with our main model end-to-end on GRID on the implementation\footnote{\url{ https://github.com/astorfi/lip-reading-deeplearning}} by \citet{3dmatch}.

For LRS2 datasets, we adopt the same structure as \citet{afouras2018compare}, which uses a 3D convolution on the input frame sequence with a filter width of 5 frames, and a 2D ResNet decreasing the spatial dimensions progressively with depth. The network convert the $T \times H \times W$ frame sequence into $T \times \frac{H}{32} \times \frac{W}{32} \times 512$ feature sequence, where $T, H, W$ is frame number, frame height, frame width respectively. It is worth noting that, training the visual front-end together with the main model could obtain poor results on LRS2, which is observed in previous works~\cite{afouras2018deepav}. Thus, as  \citet{zhao2019hearing} do, we utilize the frozen visual front-end provided by~\citet{afouras2018deepav}, which is pre-trained on a non-public datasets MV-LRS~\cite{chung2017lipprofile}, to exact the visual features. And then, we train FastLR on these features end-to-end. The pre-trained model can be found in \url{http://www.robots.ox.ac.uk/~vgg/research/deep_lip_reading/models/lrs2_lip_model.zip}.

\subsection{Model Configuration}
We adopt the Transformer~\cite{vaswani2017attention} as the basic model structure for FastLR because it is parallelizable and achieves state-of-the-art accuracy in lipreading~\cite{afouras2018deepav}. The model hidden size, number of encoder-layers, number of decoder-layers, and number of heads are set to $d_{hidden} = 512, n_{enc} = 6, n_{dec} = 6, n_{head} = 8$ for LRS2 dataset and $d_{hidden} = 256, n_{enc} = 4, n_{dec} = 4, n_{head} = 8$ for GRID dataset respectively. We replace the fully-connected network in origin Transformer with 2-layer 1D convolution network with ReLU activation which is commonly used in speech task 
and the same with TM-seq2seq~\cite{afouras2018deepav} for lipreading. The kernel size and filter size of 1D convolution are set to $4 * d_{hidden}$ and 9 respectively. The CTC decoder consists of two fully-connected layers with ReLU activation function and one fully-connected layer without activation function. The hidden sizes of these fully-connected layers equal to $d_{hidden}$. The auxiliary decoder is an ordinary Transformer decoder with the same configuration as FastLR, which takes in the target text sequence shifted right one sequence step for teacher-forcing.

\subsection{Training setup}
As mentioned in section \ref{para:encoder_enhance}, to boost the feature representation ability of encoder, we add an auxiliary connectionist temporal classification (CTC) decoder and an autoregressive decoder to FastLR and optimize them together. We set $\lambda_1$ to 0.5, $\lambda_2, \lambda_3$ to $1, 0$ during warm-up training stage, and set $\lambda_2, \lambda_3$ to $0, 1$ during main training stage for simplicity. The training steps of each training stage are listed in details in Table \ref{tab:steps}. Note that experiment on GRID dataset needs more training steps, since it is trained with its visual front-end together from scratch, different from experiments on LRS2 dataset. Moreover, the first 45k steps in warm-up stage for LRS2 are trained on LRS2-pretrain sub-dataset and all the left steps are trained on LRS2-main sub-dataset~\cite{afouras2018compare,afouras2018deepav,zhao2019hearing}.

We train our model FastLR using Adam following the optimizer settings and learning rate schedule in Transformer~\cite{vaswani2017attention}. The training procedure runs on 2 NVIDIA 1080Ti GPUs. Our code is based on tensor2tensor~\cite{tensor2tensor}.

\begin{table}[htb]
\caption{The training steps of FastLR for different datasets for each training stage.}
\centering
\begin{tabular}{l | c c }
    \toprule
    \textbf{Stage} & GRID & LRS2 \\
    \midrule
    Warm-up    & 300k & 55k \\
    Main       & 160k & 120k  \\
\bottomrule
\end{tabular}
\label{tab:steps}
\end{table}

\subsection{Inference and Evaluation}
During the inference stage, the auxiliary CTC decoder as well as autoregressive decoder will be thrown away. Given the beam size $B = 4$, FastLR generates $2*B+1$ candidates of weight embedding sequence which correspond to $2*B+1$ text sequences, and these text sequences will be sent to the decoder of a well-trained autoregressive lipreading model (TM-seq2seq) for selection as described in section \ref{sec:npd_for_cif}. The result of selected best text sequence is marked with "NPD9". We conduct the experiments on both "NPD9" and "without NPD". To be specific, the result of "without NPD" means directly using the candidate with zero-length bias without a selection process, which has a lower latency.

The recognition quality is evaluated by Word Error Rate (WER) and Character Error Rate (CER). Both error rate can be defined as:
\begin{equation}
    Error Rate = (S + D + I) / N ,
\end{equation}
where S, D, I and N are the number of substitutions, deletions, insertions and reference tokens (word or character) respectively. 

When evaluating the latency, we run FastLR on 1 NVIDIA 1080Ti GPU in inference.

\begin{table}[htb]
    \caption{The word error rate (WER) and character error rate (CER) on GRID}
    \begin{tabular}{l | c c}
      \toprule
      \multicolumn{3}{c}{GRID}  \\
      \midrule
      Method & WER & CER \\
      \midrule
      \multicolumn{1}{l}{\textit{Autoregressive Models}}         \\
      \midrule
      LSTM~\cite{wand2016lipreading} & 20.4\% & /         \\
      LipNet~\cite{assael2016lipnet} & 4.8\% & 1.9\%      \\
      WAS~\cite{chung2017lip} & \textbf{3.0\%} & /        \\
      \midrule
      \multicolumn{1}{l}{\textit{Non-Autoregressive Models}}  \\
      \midrule
      NAR-LR (base)  & 25.8\%  & 13.6\%                  \\
      FastLR (Ours)  & \textbf{4.5\%} & 2.4\%            \\
      \bottomrule
    \end{tabular}

    \label{res_grid}
\end{table}

\begin{table}[htb]
    \caption{The word error rate (WER) and character error rate (CER) on LRS2. $^\dagger$ denotes baselines from our reproduction.}
    \begin{tabular}{l | c c}
      \toprule
      \multicolumn{3}{c}{LRS2}  \\
      \midrule
      Method & WER & CER \\
      \midrule
      \multicolumn{1}{l}{\textit{Autoregressive Models}}         \\
      \midrule
      WAS~\cite{chung2017lip}  & 70.4\% & /    \\
      BLSTM+CTC~\cite{afouras2018compare} &  76.5\%    &   40.6\% \\
      LIBS~\cite{zhao2019hearing} & 65.3\% & 45.5\%         \\
      TM-seq2seq~\cite{afouras2018deepav} & \textbf{61.7\%}$^\dagger$ & 43.5\%$^\dagger$    \\
      \midrule
      \multicolumn{1}{l}{\textit{Non-Autoregressive Models}}         \\
      \midrule
      NAR-LR (base)  & 81.3\% & 57.9\%                          \\
      FastLR (Ours)  & \textbf{67.2\%} & 46.9\%                 \\
      \bottomrule
    \end{tabular}

    \label{res_lrs2}
\end{table}

\subsection{Main Results}
\label{sec:res}
We conduct experiments of FastLR, and compare them with lipreading baseline and some mainstream state-of-the-art of AR lipreading models on the GRID and LRS2 datasets respectively. As for TM-seq2seq~\cite{afouras2018deepav}, it has the same Transformer settings with FastLR and works as the AR teacher for NPD selection. We also apply CTC loss and BPE technique to TM-seq2seq for a fair comparison. \footnote{Our reproduction has a weaker performance compared with the results reported in \cite{afouras2018compare,afouras2018deepav}. Because we do not have the resource of MV-LRS, a non-public dataset which contains individual word excerpts of frequent words used by \cite{afouras2018compare, afouras2018deepav}. Thus, we do not adopt curriculum learning strategy as \citet{afouras2018compare}. }

\begin{figure}[!htb]
	\centering
	\includegraphics[width=0.46\textwidth,trim={0cm 0.0cm 0cm 0cm}, clip=true]{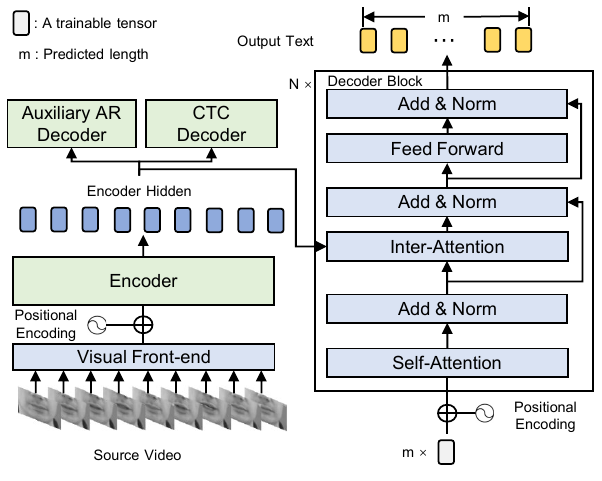}
	\caption{The NAR-LR (base) model. It is also based on Transformer~\cite{vaswani2017attention}, but generates outputs in the non-autoregressive manner~\cite{gu2017non}. It sends a series of duplicated trainable tensor into the decoder to generates target tokens. The repeat count of this trainable tensor is denoted as "m". For training, "m" is set to ground truth length, but for inference, we 
estimate it by a linear function of input length, and the parameters are obtained using the least square method on the train set. The auxiliary AR decoder is the same as FastLR's. The CTC decoder contains FC layers and CTC loss.}
	\label{nar_lr}
	\vspace{-0.2cm}
\end{figure}

The results on two datasets are listed in Table \ref{res_grid} and \ref{res_lrs2}. We can see that 1) WAS~\cite{chung2017lip} and TM-seq2seq~\cite{afouras2018compare, afouras2018deepav} obtain the best results of autoregressive lipreading model on GRID and LRS2. Compared with them, FastLR only has a slight WER absolute increase of 1.5\% and 5.5\% respectively. 2) Moreover, on GRID dataset, FastLR outperforms LipNet~\cite{assael2016lipnet} for 0.3\% WER, and exceeds LSTM~\cite{wand2016lipreading} with a notable margin; On LRS2 dataset, FastLR achieves better WER scores than WAS and BLSTM+CTC~\cite{afouras2018compare} and keeps comparable performance with LIBS~\cite{zhao2019hearing}. In addition,
compared with LIBS, we do not introduce any distillation method in training stage, and compared with WAS and TM-seq2seq, we do not leverage information from other datasets beyond GRID and LRS2.
We also propose a baseline non-autoregressive lipreading model without Integrate-and-Fire module termed as NAR-LR (base), and conduct experiments for comparison. As the result shows, FastLR outperforms this NAR baseline distinctly. The overview of the design for NAR-LR (base) is shown in Figure \ref{nar_lr}.

\subsection{Speedup}
In this section, we compare the average inference latency of FastLR with that of the autoregressive Transformer lipreading model. And then, we analyze the relationship between speedup and the length of the predicted sequence.

\subsubsection{Average Latency Comparison}
The average latency is measured in average time in seconds required to decode one sentence on the test set of LRS2 dataset. We record the inference latency and corresponding recognition accuracy of TM-seq2seq~\cite{afouras2018compare,afouras2018deepav}, FastLR without NPD and FastLR with NPD9, which is listed in Table \ref{tab:latency}. 

The result shows that FastLR speeds up the inference by 11.94$\times$ without NPD, and by 5.81$\times$ with NPD9 on average, compared with the TM-seq2seq which has similar number of model parameters. Note that the latency is calculated excluding the computation cost of data pre-processing and the visual front-end.

\begin{table}[htb]
\caption{The comparison of average inference latency and corresponding recognition accuracy. The evaluation is conducted on a server with 1 NVIDIA 1080Ti GPU, 12 Intel Xeon CPU. The batch size is set to 1. The average length of the generated sub-word sequence are all about 14.}
\centering
\begin{tabular}{l | c | c | c }
    \toprule
    \textbf{Method} & WER & Latency (s) & Speedup \\
    \midrule
    TM-seq2seq~\cite{afouras2018deepav} & 61.7\%  & 0.215 & 1.00 $\times$ \\
    FastLR (no NPD)      &  73.2\%  & 0.018  & 11.94 $\times$  \\
    FastLR (NPD 9)       &  67.2\% & 0.037 & 5.81 $\times$  \\
\bottomrule
\end{tabular}
\label{tab:latency}
\end{table}

\subsubsection{Relationship between Speedup and Length}
\begin{figure}[h]
	\centering
	
	\begin{subfigure}[h]{0.40\textwidth}
	\includegraphics[width=\textwidth]{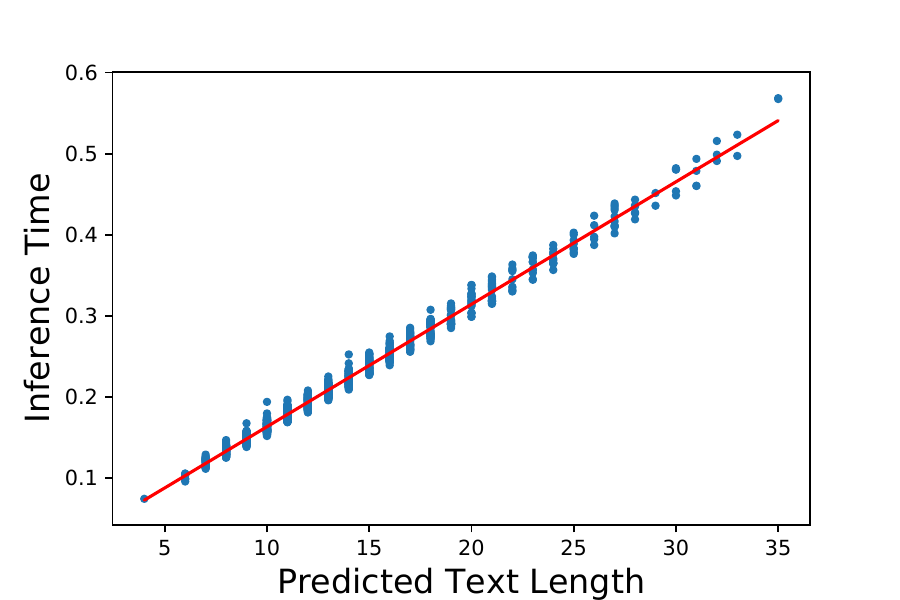}
	\caption{TM-seq2seq~\cite{afouras2018deepav}}
	\end{subfigure}
	
	\begin{subfigure}[h]{0.40\textwidth}
	\includegraphics[width=\textwidth]{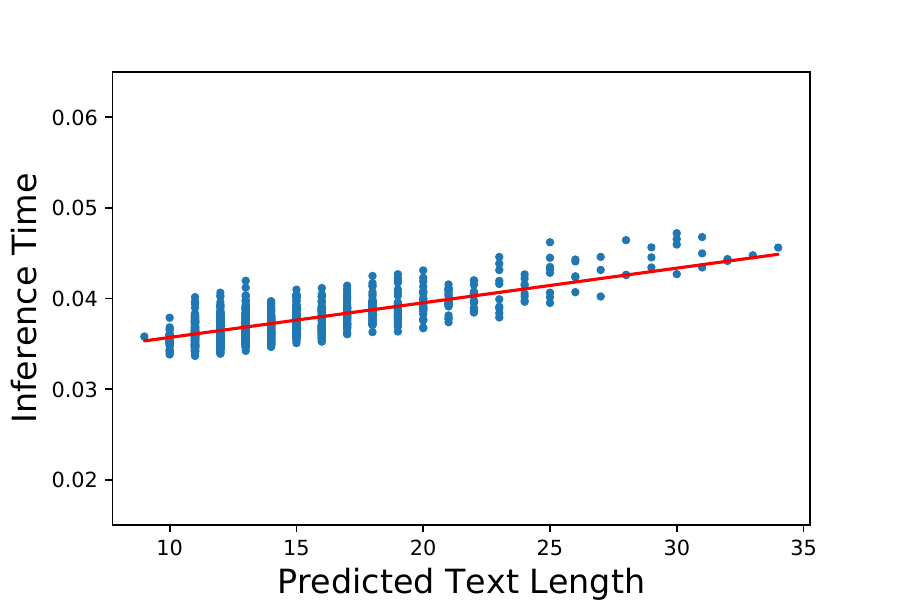}
	\caption{FastLR (NPD9)}
	\end{subfigure}
	
	\caption{Relationship between Inference time (second) and Predicted Text Length for TM-seq2seq~\cite{afouras2018deepav} and FastLR.}
	\label{seq_len_speed}
\end{figure}

During inference, the autoregressive model generates the target tokens one by one, but the non-autoregressive model speeds up the inference by increasing parallelization in the generation process. Thus, the longer the target sequence is, the more the speedup rate is. We visualize the relationship between the length of the predicted sub-word sequence in Figure \ref{seq_len_speed}. It can be seen that the inference latency increases distinctly 
with the predicted text length for TM-seq2seq, while nearly holds a  small constant for FastLR. 

Then, we bucket the test sequences of length within $[30, 35]$, and calculate their average inference latency for TM-seq2seq and FastLR to obtain the maximum speedup on LRS2 test set. The results are 0.494s and 0.045s for TM-seq2seq and FastLR (NPD9) respectively, which shows that FastLR (NPD9) achieves the speedup up to 10.97$\times$ on LRS2 test set, thanks to the parallel generation which is insensitive to sequence length.

\section{Analysis}
In this section, we first conduct ablation experiments on LRS2 to verify the significance of all proposed methods in FastLR. The experiments are listed in Table \ref{ab_study}. Then we visualize the encoder-decoder attention map of the well-trained AR model (TM-seq2seq) and the acoustic boundary detected by the I\&F module in FastLR to check whether the I\&F module works well. 

\begin{table}[!h]
    \centering
    \caption{The ablation studies on LRS2 dataset. Naive Model with I\&F is the naive lipreading model only with Integrate-and-Fire. "+Aux" means adding the auxiliary autoregressive task. We add our methods and evaluate their effectiveness progressively.}
    \begin{tabular}{ l  c  c }
        \toprule
        Model & WER & CER \\ 
    	\midrule
    	Naive Model with I\&F & >1  & 75.2\% \\
    	\midrule
    	+Aux & 93.1\% & 64.9\%  \\ 
    	+Aux+BPE & 75.7\% & 52.7\%  \\ 
    	+Aux+BPE+CTC & 73.2\% & 51.4\% \\ 
        \midrule
        +Aux+BPE+CTC+NPD \\ (FastLR) & \textbf{67.2\%} & \textbf{46.9\%} \\
    	\bottomrule
    \end{tabular}
    \label{ab_study}
    \vspace{-0.3cm}
\end{table}

\subsection{The Effectiveness of Auxiliary AR Task}
As shown in the table \ref{ab_study}, the naive lipreading model with Integrate-and-Fire is not able to converge well, due to the difficulty of learning the weight embedding in I\&F module from the meaningless encoder hidden. Thus, the autoregressive lipreading model works as the auxiliary model to enhance the feature representation ability of encoder, and guides the non-autoregressive model with Integrate-and-Fire to learn the right alignments (weight embedding). From this, the model with I\&F begins to generate the target sequence with meaning, and $CER < 65\%$ (Row 3).

\subsection{The Effectiveness of Byte-Pair Encoding}
BPE makes each token contain more language information and reduce the dependency among tokens compared with character level encoding. In addition, from observation, the speech speed of BBC video is a bit fast, which causes that one target token (character if without BPE) corresponds to few video frames. While BPE compresses the target sequence and this will help the Integrate-and-Fire module to find the acoustic level alignments easier. 

From the table \ref{ab_study} (Row 4), it can be seen that BPE reduces the word error rate and character error rate to 75.7\% and 52.7\% respectively, which means BPE helps the model gains the ability to generates understandable  sentence. 

\subsection{The Effectiveness of CTC}
The result shows that (Row 5), adding auxiliary connectionist temporal classification(CTC) decoder with CTC loss will further boost the feature representation ability of encoder, and cause 2.5\%  absolute decrease in WER. At this point, the model gains considerable recognition accuracy compared with the traditional autoregressive method. 

\subsection{The Effectiveness of NPD for I\&F}
\begin{figure}[!htb]
	\centering
	\includegraphics[width=0.46\textwidth,trim={0cm 0.0cm 0cm 0cm}, clip=true]{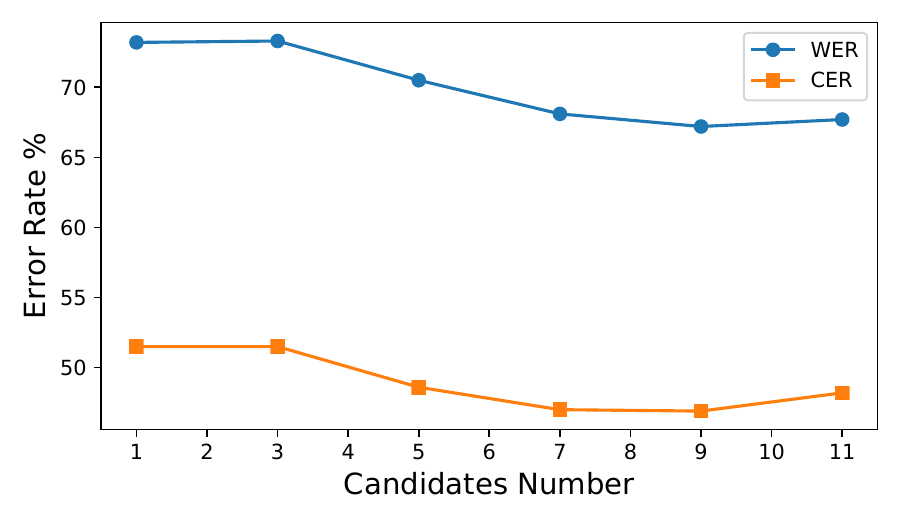}
	\caption{The effect of cadidates number on WER and CER for FastLR model.}
	\label{cad_num}
	\vspace{-0.2cm}
\end{figure}

Table \ref{ab_study} (Row 6) shows that using NPD for I\&F can boost the performance effectively. We also study the effect of increasing the candidates number for FastLR on LRS2 dataset, as shown in Figure \ref{cad_num}. It can be seen that, when setting the candidates number to $9$, the accuracy peaks. Finally, FastLR achieves considerable accuracy compared with state-of-the-art autoregressive lipreading model.

\subsection{The Visualization of Boundary Detection}
\begin{figure}[htb]
	\centering
	\includegraphics[width=0.48\textwidth,trim={0cm 0.0cm 0cm 0cm}, clip=true]{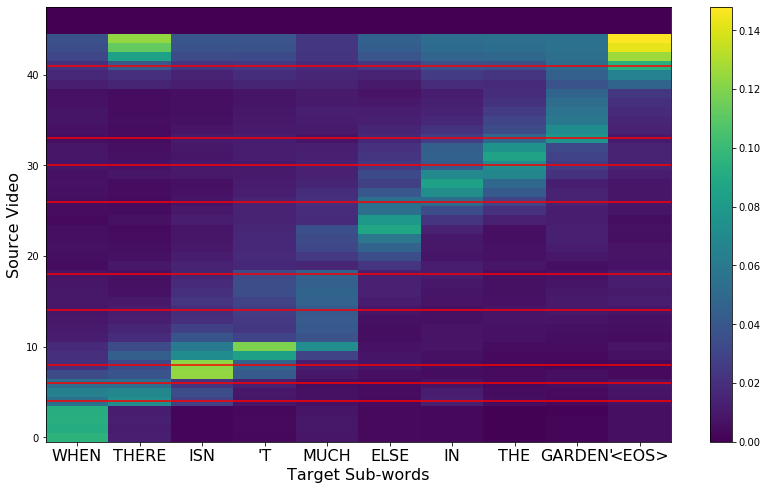}
	\caption{An example of the visualization for encoder-decoder attention map and the acoustic boundary. The horizontal red lines represent the acoustic boundaries detected by I\&F module in FastLR, which split the video frames to discrete segments.}
	\label{fig:alignments}
\end{figure}

We visualize the encoder-decoder attention map in Figure \ref{fig:alignments}, which is obtained from the well-trained AR TM-seq2seq. The attention map illustrates the alignment between source video frames and the corresponding target sub-word sequence. 

The figure shows that the video frames between two horizontal red lines are roughly just what the corresponding target token attends to. It means that the "accumulate and detect" part in I\&F module tells the acoustic boundary well and makes a right prediction of sequence length.

\section{Conclusion}
In this work, we developed FastLR, a non-autoregressive lipreading system with Integrate-and-Fire module, that recognizes source silent video and generates all the target text tokens in parallel. FastLR consists of a visual front-end, a visual feature encoder and a text decoder for simultaneous generation. To bridge the accuracy gap between FastLR and state-of-the-art autoregressive lipreading model, we introduce I\&F module to encode the continuous visual features into discrete token embedding by locating the acoustic boundary. In addition, we propose several methods including auxiliary AR task and CTC loss to boost the feature representation ability of encoder. At last, we design NPD for I\&F and bring Byte-Pair Encoding into lipreading, and both methods alleviate the problem caused by the removal of AR language model. Experiments on GRID and LRS2 lipreading datasets show that FastLR outperforms the NAR-LR baseline and has a slight WER increase compared with state-of-the-art AR model, which demonstrates the effectiveness of our method for NAR lipreading.

In the future, we will continue to work on how to make a better approximation to the true target distribution for NAR lipreading task, and design more flexible policies to bridge the gap between AR and NAR model as well as keeping the fast speed of NAR generation.

\section*{Acknowledgments}
This work was supported in part by the National Key R\&D Program of China (Grant No.2018AAA0100603), Zhejiang Natural Science Foundation (LR19F020006), National Natural Science Foundation of China (Grant No.61836002, No.U1611461 and No.61751209) and the Fundamental Research Funds for the Central Universities (2020QNA5024). This work was also partially supported by the Language and Speech Innovation Lab of HUAWEI Cloud.

\newpage
\bibliographystyle{ACM-Reference-Format}
\balance 
\bibliography{sample-base}


\begin{thebibliography}{34}


\ifx \showCODEN    \undefined \def \showCODEN     #1{\unskip}     \fi
\ifx \showDOI      \undefined \def \showDOI       #1{#1}\fi
\ifx \showISBNx    \undefined \def \showISBNx     #1{\unskip}     \fi
\ifx \showISBNxiii \undefined \def \showISBNxiii  #1{\unskip}     \fi
\ifx \showISSN     \undefined \def \showISSN      #1{\unskip}     \fi
\ifx \showLCCN     \undefined \def \showLCCN      #1{\unskip}     \fi
\ifx \shownote     \undefined \def \shownote      #1{#1}          \fi
\ifx \showarticletitle \undefined \def \showarticletitle #1{#1}   \fi
\ifx \showURL      \undefined \def \showURL       {\relax}        \fi
\providecommand\bibfield[2]{#2}
\providecommand\bibinfo[2]{#2}
\providecommand\natexlab[1]{#1}
\providecommand\showeprint[2][]{arXiv:#2}

\bibitem[\protect\citeauthoryear{Afouras, Chung, Senior, Vinyals, and
  Zisserman}{Afouras et~al\mbox{.}}{2018b}]%
        {afouras2018deepav}
\bibfield{author}{\bibinfo{person}{Triantafyllos Afouras},
  \bibinfo{person}{Joon~Son Chung}, \bibinfo{person}{Andrew Senior},
  \bibinfo{person}{Oriol Vinyals}, {and} \bibinfo{person}{Andrew Zisserman}.}
  \bibinfo{year}{2018}\natexlab{b}.
\newblock \showarticletitle{Deep audio-visual speech recognition}.
\newblock \bibinfo{journal}{\emph{IEEE transactions on pattern analysis and
  machine intelligence}} (\bibinfo{year}{2018}).
\newblock


\bibitem[\protect\citeauthoryear{Afouras, Chung, and Zisserman}{Afouras
  et~al\mbox{.}}{2018a}]%
        {afouras2018compare}
\bibfield{author}{\bibinfo{person}{Triantafyllos Afouras},
  \bibinfo{person}{Joon~Son Chung}, {and} \bibinfo{person}{Andrew Zisserman}.}
  \bibinfo{year}{2018}\natexlab{a}.
\newblock \showarticletitle{Deep lip reading: a comparison of models and an
  online application}.
\newblock \bibinfo{journal}{\emph{arXiv preprint arXiv:1806.06053}}
  (\bibinfo{year}{2018}).
\newblock


\bibitem[\protect\citeauthoryear{Assael, Shillingford, Whiteson, and
  De~Freitas}{Assael et~al\mbox{.}}{2016}]%
        {assael2016lipnet}
\bibfield{author}{\bibinfo{person}{Yannis~M Assael}, \bibinfo{person}{Brendan
  Shillingford}, \bibinfo{person}{Shimon Whiteson}, {and}
  \bibinfo{person}{Nando De~Freitas}.} \bibinfo{year}{2016}\natexlab{}.
\newblock \showarticletitle{Lipnet: End-to-end sentence-level lipreading}.
\newblock \bibinfo{journal}{\emph{arXiv preprint arXiv:1611.01599}}
  (\bibinfo{year}{2016}).
\newblock


\bibitem[\protect\citeauthoryear{Burkitt}{Burkitt}{2006}]%
        {burkitt2006review}
\bibfield{author}{\bibinfo{person}{Anthony~N Burkitt}.}
  \bibinfo{year}{2006}\natexlab{}.
\newblock \showarticletitle{A review of the integrate-and-fire neuron model: I.
  Homogeneous synaptic input}.
\newblock \bibinfo{journal}{\emph{Biological cybernetics}}
  \bibinfo{volume}{95}, \bibinfo{number}{1} (\bibinfo{year}{2006}),
  \bibinfo{pages}{1--19}.
\newblock


\bibitem[\protect\citeauthoryear{Chen, Watanabe, Villalba, and Dehak}{Chen
  et~al\mbox{.}}{2019}]%
        {chen2019non}
\bibfield{author}{\bibinfo{person}{Nanxin Chen}, \bibinfo{person}{Shinji
  Watanabe}, \bibinfo{person}{Jes{\'u}s Villalba}, {and} \bibinfo{person}{Najim
  Dehak}.} \bibinfo{year}{2019}\natexlab{}.
\newblock \showarticletitle{Non-Autoregressive Transformer Automatic Speech
  Recognition}.
\newblock \bibinfo{journal}{\emph{arXiv preprint arXiv:1911.04908}}
  (\bibinfo{year}{2019}).
\newblock


\bibitem[\protect\citeauthoryear{Chung, Gulcehre, Cho, and Bengio}{Chung
  et~al\mbox{.}}{2014}]%
        {chung2014gru}
\bibfield{author}{\bibinfo{person}{Junyoung Chung}, \bibinfo{person}{Caglar
  Gulcehre}, \bibinfo{person}{KyungHyun Cho}, {and} \bibinfo{person}{Yoshua
  Bengio}.} \bibinfo{year}{2014}\natexlab{}.
\newblock \showarticletitle{Empirical evaluation of gated recurrent neural
  networks on sequence modeling}.
\newblock \bibinfo{journal}{\emph{arXiv preprint arXiv:1412.3555}}
  (\bibinfo{year}{2014}).
\newblock


\bibitem[\protect\citeauthoryear{Chung, Senior, Vinyals, and Zisserman}{Chung
  et~al\mbox{.}}{2017}]%
        {chung2017lip}
\bibfield{author}{\bibinfo{person}{Joon~Son Chung}, \bibinfo{person}{Andrew
  Senior}, \bibinfo{person}{Oriol Vinyals}, {and} \bibinfo{person}{Andrew
  Zisserman}.} \bibinfo{year}{2017}\natexlab{}.
\newblock \showarticletitle{Lip reading sentences in the wild}. In
  \bibinfo{booktitle}{\emph{2017 IEEE Conference on Computer Vision and Pattern
  Recognition (CVPR)}}. IEEE, \bibinfo{pages}{3444--3453}.
\newblock


\bibitem[\protect\citeauthoryear{Chung and Zisserman}{Chung and
  Zisserman}{2017}]%
        {chung2017lipprofile}
\bibfield{author}{\bibinfo{person}{Joon~Son Chung} {and} \bibinfo{person}{AP
  Zisserman}.} \bibinfo{year}{2017}\natexlab{}.
\newblock \showarticletitle{Lip reading in profile}.
\newblock  (\bibinfo{year}{2017}).
\newblock


\bibitem[\protect\citeauthoryear{Cooke, Barker, Cunningham, and Shao}{Cooke
  et~al\mbox{.}}{2006}]%
        {griddataset}
\bibfield{author}{\bibinfo{person}{Martin Cooke}, \bibinfo{person}{Jon Barker},
  \bibinfo{person}{Stuart Cunningham}, {and} \bibinfo{person}{Xu Shao}.}
  \bibinfo{year}{2006}\natexlab{}.
\newblock \showarticletitle{An audio-visual corpus for speech perception and
  automatic speech recognition}.
\newblock \bibinfo{journal}{\emph{The Journal of the Acoustical Society of
  America}} \bibinfo{volume}{120}, \bibinfo{number}{5} (\bibinfo{year}{2006}),
  \bibinfo{pages}{2421--2424}.
\newblock


\bibitem[\protect\citeauthoryear{Dong and Xu}{Dong and Xu}{2019}]%
        {dong2019cif}
\bibfield{author}{\bibinfo{person}{Linhao Dong} {and} \bibinfo{person}{Bo Xu}.}
  \bibinfo{year}{2019}\natexlab{}.
\newblock \showarticletitle{CIF: Continuous Integrate-and-Fire for End-to-End
  Speech Recognition}.
\newblock \bibinfo{journal}{\emph{arXiv preprint arXiv:1905.11235}}
  (\bibinfo{year}{2019}).
\newblock


\bibitem[\protect\citeauthoryear{Ghazvininejad, Levy, Liu, and
  Zettlemoyer}{Ghazvininejad et~al\mbox{.}}{2019}]%
        {ghazvininejad2019mask}
\bibfield{author}{\bibinfo{person}{Marjan Ghazvininejad}, \bibinfo{person}{Omer
  Levy}, \bibinfo{person}{Yinhan Liu}, {and} \bibinfo{person}{Luke
  Zettlemoyer}.} \bibinfo{year}{2019}\natexlab{}.
\newblock \showarticletitle{Mask-predict: Parallel decoding of conditional
  masked language models}. In \bibinfo{booktitle}{\emph{Proceedings of the 2019
  Conference on Empirical Methods in Natural Language Processing and the 9th
  International Joint Conference on Natural Language Processing
  (EMNLP-IJCNLP)}}. \bibinfo{pages}{6114--6123}.
\newblock


\bibitem[\protect\citeauthoryear{Graves, Fern{\'a}ndez, Gomez, and
  Schmidhuber}{Graves et~al\mbox{.}}{2006}]%
        {graves2006ctc}
\bibfield{author}{\bibinfo{person}{Alex Graves}, \bibinfo{person}{Santiago
  Fern{\'a}ndez}, \bibinfo{person}{Faustino Gomez}, {and}
  \bibinfo{person}{J{\"u}rgen Schmidhuber}.} \bibinfo{year}{2006}\natexlab{}.
\newblock \showarticletitle{Connectionist temporal classification: labelling
  unsegmented sequence data with recurrent neural networks}. In
  \bibinfo{booktitle}{\emph{Proceedings of the 23rd international conference on
  Machine learning}}. \bibinfo{pages}{369--376}.
\newblock


\bibitem[\protect\citeauthoryear{Gu, Bradbury, Xiong, Li, and Socher}{Gu
  et~al\mbox{.}}{2017}]%
        {gu2017non}
\bibfield{author}{\bibinfo{person}{Jiatao Gu}, \bibinfo{person}{James
  Bradbury}, \bibinfo{person}{Caiming Xiong}, \bibinfo{person}{Victor~OK Li},
  {and} \bibinfo{person}{Richard Socher}.} \bibinfo{year}{2017}\natexlab{}.
\newblock \showarticletitle{Non-autoregressive neural machine translation}.
\newblock \bibinfo{journal}{\emph{arXiv preprint arXiv:1711.02281}}
  (\bibinfo{year}{2017}).
\newblock


\bibitem[\protect\citeauthoryear{Guo, Tan, He, Qin, Xu, and Liu}{Guo
  et~al\mbox{.}}{2019}]%
        {guo2019non}
\bibfield{author}{\bibinfo{person}{Junliang Guo}, \bibinfo{person}{Xu Tan},
  \bibinfo{person}{Di He}, \bibinfo{person}{Tao Qin}, \bibinfo{person}{Linli
  Xu}, {and} \bibinfo{person}{Tie-Yan Liu}.} \bibinfo{year}{2019}\natexlab{}.
\newblock \showarticletitle{Non-autoregressive neural machine translation with
  enhanced decoder input}. In \bibinfo{booktitle}{\emph{AAAI}},
  Vol.~\bibinfo{volume}{33}. \bibinfo{pages}{3723--3730}.
\newblock


\bibitem[\protect\citeauthoryear{Lee, Mansimov, and Cho}{Lee
  et~al\mbox{.}}{2018}]%
        {lee2018deterministic}
\bibfield{author}{\bibinfo{person}{Jason Lee}, \bibinfo{person}{Elman
  Mansimov}, {and} \bibinfo{person}{Kyunghyun Cho}.}
  \bibinfo{year}{2018}\natexlab{}.
\newblock \showarticletitle{Deterministic Non-Autoregressive Neural Sequence
  Modeling by Iterative Refinement}. In \bibinfo{booktitle}{\emph{EMNLP}}.
  \bibinfo{pages}{1173--1182}.
\newblock


\bibitem[\protect\citeauthoryear{Liu, Ren, Tan, Zhang, Qin, Zhao, and Liu}{Liu
  et~al\mbox{.}}{2020}]%
        {liu2020task}
\bibfield{author}{\bibinfo{person}{Jinglin Liu}, \bibinfo{person}{Yi Ren},
  \bibinfo{person}{Xu Tan}, \bibinfo{person}{Chen Zhang}, \bibinfo{person}{Tao
  Qin}, \bibinfo{person}{Zhou Zhao}, {and} \bibinfo{person}{Tie-Yan Liu}.}
  \bibinfo{year}{2020}\natexlab{}.
\newblock \showarticletitle{Task-Level Curriculum Learning for
  Non-Autoregressive Neural Machine Translation}. In
  \bibinfo{booktitle}{\emph{Proceedings of the Twenty-Ninth International Joint
  Conference on Artificial Intelligence, {IJCAI-20}}}.
  \bibinfo{pages}{3861--3867}.
\newblock


\bibitem[\protect\citeauthoryear{Ma, Zhou, Li, Neubig, and Hovy}{Ma
  et~al\mbox{.}}{2019}]%
        {ma2019flowseq}
\bibfield{author}{\bibinfo{person}{Xuezhe Ma}, \bibinfo{person}{Chunting Zhou},
  \bibinfo{person}{Xian Li}, \bibinfo{person}{Graham Neubig}, {and}
  \bibinfo{person}{Eduard Hovy}.} \bibinfo{year}{2019}\natexlab{}.
\newblock \showarticletitle{FlowSeq: Non-Autoregressive Conditional Sequence
  Generation with Generative Flow}. In
  \bibinfo{booktitle}{\emph{EMNLP-IJCNLP}}. \bibinfo{pages}{4273--4283}.
\newblock


\bibitem[\protect\citeauthoryear{Oord, Li, Babuschkin, Simonyan, Vinyals,
  Kavukcuoglu, Driessche, Lockhart, Cobo, Stimberg, et~al\mbox{.}}{Oord
  et~al\mbox{.}}{2017}]%
        {oord2017parallel}
\bibfield{author}{\bibinfo{person}{Aaron van~den Oord}, \bibinfo{person}{Yazhe
  Li}, \bibinfo{person}{Igor Babuschkin}, \bibinfo{person}{Karen Simonyan},
  \bibinfo{person}{Oriol Vinyals}, \bibinfo{person}{Koray Kavukcuoglu},
  \bibinfo{person}{George van~den Driessche}, \bibinfo{person}{Edward
  Lockhart}, \bibinfo{person}{Luis~C Cobo}, \bibinfo{person}{Florian Stimberg},
  {et~al\mbox{.}}} \bibinfo{year}{2017}\natexlab{}.
\newblock \showarticletitle{Parallel wavenet: Fast high-fidelity speech
  synthesis}.
\newblock \bibinfo{journal}{\emph{arXiv preprint arXiv:1711.10433}}
  (\bibinfo{year}{2017}).
\newblock


\bibitem[\protect\citeauthoryear{Petridis, Stafylakis, Ma, Tzimiropoulos, and
  Pantic}{Petridis et~al\mbox{.}}{2018}]%
        {ctchyp2018}
\bibfield{author}{\bibinfo{person}{Stavros Petridis}, \bibinfo{person}{Themos
  Stafylakis}, \bibinfo{person}{Pingchuan Ma}, \bibinfo{person}{Georgios
  Tzimiropoulos}, {and} \bibinfo{person}{Maja Pantic}.}
  \bibinfo{year}{2018}\natexlab{}.
\newblock \showarticletitle{Audio-visual speech recognition with a hybrid
  ctc/attention architecture}. In \bibinfo{booktitle}{\emph{2018 IEEE Spoken
  Language Technology Workshop (SLT)}}. IEEE, \bibinfo{pages}{513--520}.
\newblock


\bibitem[\protect\citeauthoryear{Ren, Hu, Qin, Zhao, Zhao, and Liu}{Ren
  et~al\mbox{.}}{2020a}]%
        {ren2020fastspeech}
\bibfield{author}{\bibinfo{person}{Yi Ren}, \bibinfo{person}{Chenxu Hu},
  \bibinfo{person}{Tao Qin}, \bibinfo{person}{Sheng Zhao},
  \bibinfo{person}{Zhou Zhao}, {and} \bibinfo{person}{Tie-Yan Liu}.}
  \bibinfo{year}{2020}\natexlab{a}.
\newblock \showarticletitle{FastSpeech 2: Fast and High-Quality End-to-End
  Text-to-Speech}.
\newblock \bibinfo{journal}{\emph{arXiv preprint arXiv:2006.04558}}
  (\bibinfo{year}{2020}).
\newblock


\bibitem[\protect\citeauthoryear{Ren, Liu, Tan, Zhao, Zhao, and Liu}{Ren
  et~al\mbox{.}}{2020b}]%
        {ren2020study}
\bibfield{author}{\bibinfo{person}{Yi Ren}, \bibinfo{person}{Jinglin Liu},
  \bibinfo{person}{Xu Tan}, \bibinfo{person}{Sheng Zhao}, \bibinfo{person}{Zhou
  Zhao}, {and} \bibinfo{person}{Tie-Yan Liu}.}
  \bibinfo{year}{2020}\natexlab{b}.
\newblock \showarticletitle{A Study of Non-autoregressive Model for Sequence
  Generation}.
\newblock \bibinfo{journal}{\emph{arXiv preprint arXiv:2004.10454}}
  (\bibinfo{year}{2020}).
\newblock


\bibitem[\protect\citeauthoryear{Ren, Ruan, Tan, Qin, Zhao, Zhao, and Liu}{Ren
  et~al\mbox{.}}{2019}]%
        {ren2019fastspeech}
\bibfield{author}{\bibinfo{person}{Yi Ren}, \bibinfo{person}{Yangjun Ruan},
  \bibinfo{person}{Xu Tan}, \bibinfo{person}{Tao Qin}, \bibinfo{person}{Sheng
  Zhao}, \bibinfo{person}{Zhou Zhao}, {and} \bibinfo{person}{Tie-Yan Liu}.}
  \bibinfo{year}{2019}\natexlab{}.
\newblock \showarticletitle{Fastspeech: Fast, robust and controllable text to
  speech}. In \bibinfo{booktitle}{\emph{Advances in Neural Information
  Processing Systems}}. \bibinfo{pages}{3165--3174}.
\newblock


\bibitem[\protect\citeauthoryear{Sennrich, Haddow, and Birch}{Sennrich
  et~al\mbox{.}}{2015}]%
        {sennrich2015neural}
\bibfield{author}{\bibinfo{person}{Rico Sennrich}, \bibinfo{person}{Barry
  Haddow}, {and} \bibinfo{person}{Alexandra Birch}.}
  \bibinfo{year}{2015}\natexlab{}.
\newblock \showarticletitle{Neural machine translation of rare words with
  subword units}.
\newblock \bibinfo{journal}{\emph{arXiv preprint arXiv:1508.07909}}
  (\bibinfo{year}{2015}).
\newblock


\bibitem[\protect\citeauthoryear{Shillingford, Assael, Hoffman, Paine, Hughes,
  Prabhu, Liao, Sak, Rao, Bennett, et~al\mbox{.}}{Shillingford
  et~al\mbox{.}}{2018}]%
        {shillingford2018large}
\bibfield{author}{\bibinfo{person}{Brendan Shillingford},
  \bibinfo{person}{Yannis Assael}, \bibinfo{person}{Matthew~W Hoffman},
  \bibinfo{person}{Thomas Paine}, \bibinfo{person}{C{\'\i}an Hughes},
  \bibinfo{person}{Utsav Prabhu}, \bibinfo{person}{Hank Liao},
  \bibinfo{person}{Hasim Sak}, \bibinfo{person}{Kanishka Rao},
  \bibinfo{person}{Lorrayne Bennett}, {et~al\mbox{.}}}
  \bibinfo{year}{2018}\natexlab{}.
\newblock \showarticletitle{Large-scale visual speech recognition}.
\newblock \bibinfo{journal}{\emph{arXiv preprint arXiv:1807.05162}}
  (\bibinfo{year}{2018}).
\newblock


\bibitem[\protect\citeauthoryear{Stafylakis and Tzimiropoulos}{Stafylakis and
  Tzimiropoulos}{2017}]%
        {stafylakis2017combining}
\bibfield{author}{\bibinfo{person}{Themos Stafylakis} {and}
  \bibinfo{person}{Georgios Tzimiropoulos}.} \bibinfo{year}{2017}\natexlab{}.
\newblock \showarticletitle{Combining residual networks with LSTMs for
  lipreading}.
\newblock \bibinfo{journal}{\emph{arXiv preprint arXiv:1703.04105}}
  (\bibinfo{year}{2017}).
\newblock


\bibitem[\protect\citeauthoryear{Sutskever, Vinyals, and Le}{Sutskever
  et~al\mbox{.}}{2014}]%
        {sutskever2014sequence}
\bibfield{author}{\bibinfo{person}{Ilya Sutskever}, \bibinfo{person}{Oriol
  Vinyals}, {and} \bibinfo{person}{Quoc~V Le}.}
  \bibinfo{year}{2014}\natexlab{}.
\newblock \showarticletitle{Sequence to sequence learning with neural
  networks}. In \bibinfo{booktitle}{\emph{Advances in neural information
  processing systems}}. \bibinfo{pages}{3104--3112}.
\newblock


\bibitem[\protect\citeauthoryear{Torfi, Iranmanesh, Nasrabadi, and
  Dawson}{Torfi et~al\mbox{.}}{2017}]%
        {3dmatch}
\bibfield{author}{\bibinfo{person}{Amirsina Torfi},
  \bibinfo{person}{Seyed~Mehdi Iranmanesh}, \bibinfo{person}{Nasser Nasrabadi},
  {and} \bibinfo{person}{Jeremy Dawson}.} \bibinfo{year}{2017}\natexlab{}.
\newblock \showarticletitle{3d convolutional neural networks for cross
  audio-visual matching recognition}.
\newblock \bibinfo{journal}{\emph{IEEE Access}}  \bibinfo{volume}{5}
  (\bibinfo{year}{2017}), \bibinfo{pages}{22081--22091}.
\newblock


\bibitem[\protect\citeauthoryear{Vaswani, Bengio, Brevdo, Chollet, Gomez,
  Gouws, Jones, Kaiser, Kalchbrenner, Parmar, Sepassi, Shazeer, and
  Uszkoreit}{Vaswani et~al\mbox{.}}{2018}]%
        {tensor2tensor}
\bibfield{author}{\bibinfo{person}{Ashish Vaswani}, \bibinfo{person}{Samy
  Bengio}, \bibinfo{person}{Eugene Brevdo}, \bibinfo{person}{Francois Chollet},
  \bibinfo{person}{Aidan~N. Gomez}, \bibinfo{person}{Stephan Gouws},
  \bibinfo{person}{Llion Jones}, \bibinfo{person}{\L{}ukasz Kaiser},
  \bibinfo{person}{Nal Kalchbrenner}, \bibinfo{person}{Niki Parmar},
  \bibinfo{person}{Ryan Sepassi}, \bibinfo{person}{Noam Shazeer}, {and}
  \bibinfo{person}{Jakob Uszkoreit}.} \bibinfo{year}{2018}\natexlab{}.
\newblock \showarticletitle{Tensor2Tensor for Neural Machine Translation}.
\newblock \bibinfo{journal}{\emph{CoRR}}  \bibinfo{volume}{abs/1803.07416}
  (\bibinfo{year}{2018}).
\newblock
\urldef\tempurl%
\url{http://arxiv.org/abs/1803.07416}
\showURL{%
\tempurl}


\bibitem[\protect\citeauthoryear{Vaswani, Shazeer, Parmar, Uszkoreit, Jones,
  Gomez, Kaiser, and Polosukhin}{Vaswani et~al\mbox{.}}{2017}]%
        {vaswani2017attention}
\bibfield{author}{\bibinfo{person}{Ashish Vaswani}, \bibinfo{person}{Noam
  Shazeer}, \bibinfo{person}{Niki Parmar}, \bibinfo{person}{Jakob Uszkoreit},
  \bibinfo{person}{Llion Jones}, \bibinfo{person}{Aidan~N Gomez},
  \bibinfo{person}{{\L}ukasz Kaiser}, {and} \bibinfo{person}{Illia
  Polosukhin}.} \bibinfo{year}{2017}\natexlab{}.
\newblock \showarticletitle{Attention is all you need}. In
  \bibinfo{booktitle}{\emph{Advances in neural information processing
  systems}}. \bibinfo{pages}{5998--6008}.
\newblock


\bibitem[\protect\citeauthoryear{Wand, Koutn{\'\i}k, and Schmidhuber}{Wand
  et~al\mbox{.}}{2016}]%
        {wand2016lipreading}
\bibfield{author}{\bibinfo{person}{Michael Wand}, \bibinfo{person}{Jan
  Koutn{\'\i}k}, {and} \bibinfo{person}{J{\"u}rgen Schmidhuber}.}
  \bibinfo{year}{2016}\natexlab{}.
\newblock \showarticletitle{Lipreading with long short-term memory}. In
  \bibinfo{booktitle}{\emph{2016 IEEE International Conference on Acoustics,
  Speech and Signal Processing (ICASSP)}}. IEEE, \bibinfo{pages}{6115--6119}.
\newblock


\bibitem[\protect\citeauthoryear{Wang, Tian, He, Qin, Zhai, and Liu}{Wang
  et~al\mbox{.}}{2019}]%
        {wang2019non}
\bibfield{author}{\bibinfo{person}{Yiren Wang}, \bibinfo{person}{Fei Tian},
  \bibinfo{person}{Di He}, \bibinfo{person}{Tao Qin},
  \bibinfo{person}{ChengXiang Zhai}, {and} \bibinfo{person}{Tie-Yan Liu}.}
  \bibinfo{year}{2019}\natexlab{}.
\newblock \showarticletitle{Non-Autoregressive Machine Translation with
  Auxiliary Regularization}. In \bibinfo{booktitle}{\emph{AAAI}}.
\newblock


\bibitem[\protect\citeauthoryear{Yang, Liu, and Zou}{Yang
  et~al\mbox{.}}{2019}]%
        {yang2019non}
\bibfield{author}{\bibinfo{person}{Bang Yang}, \bibinfo{person}{Fenglin Liu},
  {and} \bibinfo{person}{Yuexian Zou}.} \bibinfo{year}{2019}\natexlab{}.
\newblock \showarticletitle{Non-Autoregressive Video Captioning with Iterative
  Refinement}.
\newblock \bibinfo{journal}{\emph{arXiv preprint arXiv:1911.12018}}
  (\bibinfo{year}{2019}).
\newblock


\bibitem[\protect\citeauthoryear{Zhang, Cheng, and Wang}{Zhang
  et~al\mbox{.}}{2019}]%
        {zhang2019spatio}
\bibfield{author}{\bibinfo{person}{Xingxuan Zhang}, \bibinfo{person}{Feng
  Cheng}, {and} \bibinfo{person}{Shilin Wang}.}
  \bibinfo{year}{2019}\natexlab{}.
\newblock \showarticletitle{Spatio-temporal fusion based convolutional sequence
  learning for lip reading}. In \bibinfo{booktitle}{\emph{Proceedings of the
  IEEE/CVF International Conference on Computer Vision}}.
  \bibinfo{pages}{713--722}.
\newblock


\bibitem[\protect\citeauthoryear{Zhao, Xu, Wang, Hou, Tang, and Song}{Zhao
  et~al\mbox{.}}{2019}]%
        {zhao2019hearing}
\bibfield{author}{\bibinfo{person}{Ya Zhao}, \bibinfo{person}{Rui Xu},
  \bibinfo{person}{Xinchao Wang}, \bibinfo{person}{Peng Hou},
  \bibinfo{person}{Haihong Tang}, {and} \bibinfo{person}{Mingli Song}.}
  \bibinfo{year}{2019}\natexlab{}.
\newblock \showarticletitle{Hearing Lips: Improving Lip Reading by Distilling
  Speech Recognizers}.
\newblock \bibinfo{journal}{\emph{arXiv preprint arXiv:1911.11502}}
  (\bibinfo{year}{2019}).
\newblock


\end{thebibliography}

\end{document}